\documentclass[letters,fleqn,usenatbib]{mnras}

\usepackage[T1]{fontenc}

\DeclareRobustCommand{\VAN}[3]{#2}
\let\VANthebibliography\thebibliography
\def\thebibliography{\DeclareRobustCommand{\VAN}[3]{##3}\VANthebibliography}

\usepackage{graphicx}	
\usepackage{amsmath}	
\usepackage{amssymb}	

\newcommand{\LX}{$\left<{\rm L}_{\rm X}\right>$}
\newcommand{\LXMS}{$\left<{\rm L}_{\rm X}\right>-{\rm M}_*$}
\newcommand{\MBHMS}{M$_{\rm BH}-{\rm M}_*$}
\newcommand{\PLz}{$P(\log \lambda,z)$}
\newcommand{\UMBHz}{$U(\log {\rm M}_{\rm BH},z)$}

\title[The origin of the ${\rm L}_{\rm X}-{\rm M}_{*}$ relation]{An Eddington ratio-driven origin for the ${\rm L}_{\rm X}-{\rm M}_{*}$ relation in quiescent and star forming active galaxies}
\author[R. Carraro et al.]{
Rosamaria Carraro,$^{1}$\thanks{E-mail: rosamaria.carraro@postgrado.uv.cl}
Francesco Shankar,$^{2}$\thanks{E-mail: F.Shankar@soton.ac.uk}
Viola Allevato,$^{3,4}$
Giulia Rodighiero,$^{5}$
Christopher Marsden,$^{2}$
\newauthor Patricia Ar\'evalo,$^{1}$
Ivan Delvecchio,$^{6}$
Andrea Lapi$^{7,8,9,10}$
\\
$^{1}$Instituto de F\'\i{}sica y Astronom\'\i{}a, Universidad de Valpara\'\i{}so, Gran Bretaña 1111, Playa Ancha, Valpara\'\i{}so, Chile\\
$^{2}$School of Physics and Astronomy, University of Southampton, Highfield SO17 1BJ, UK\\
$^{3}$INAF-Osservatorio di Astrofisica e Scienza dello Spazio di Bologna, I-40129 Bologna, Italy\\
$^{4}$Scuola Normale Superiore, Piazza dei Cavalieri 7, I-56126 Pisa, Italy\\
$^{5}$Dipartimento di Fisica e Astronomia, Universit\`a di Padova, Vicolo dell'Osservatorio, 3, I-35122, Padova, Italy\\
$^{6}$INAF - Osservatorio Astronomico di Brera, via Brera 28, I-20121, Milano, Italy \& via Bianchi 46, I-23807, Merate, Italy\\
$^{7}$SISSA, Via Bonomea 265, 34136 Trieste, Italy\\
$^{8}$IFPU - Institute for fundamental physics of the Universe, Via Beirut 2, 34014 Trieste, Italy\\
$^{9}$INFN-Sezione di Trieste, via Valerio 2, 34127 Trieste,  Italy\\
$^{10}$INAF-Osservatorio Astronomico di Trieste, via Tiepolo 11, 34131 Trieste, Italy\\
}

\date{Accepted XXX. Received YYY; in original form ZZZ}

\pubyear{2021}

\usepackage{newtxtext,newtxmath}

\begin{document}
\label{firstpage}
\pagerange{\pageref{firstpage}--\pageref{lastpage}}
\maketitle

\begin{abstract} 
A mild correlation exists in active galaxies between the mean black hole accretion, as traced by the mean X-ray luminosity \LX\, and the host galaxy stellar mass M$_*$, characterised by a normalisation steadily decreasing with cosmic time and lower in more quiescent galaxies. 
We create comprehensive semi-empirical mock catalogues of active black holes to pin down which parameters control the shape and evolution of the \LXMS\ relation of X-ray detected active galaxies. 
We find that the normalisation of the \LXMS\ relation is largely independent of the fraction of active galaxies (the duty cycle), but strongly dependent on the mean Eddington ratio, when adopting a constant underlying \MBHMS\ relation as suggested by observational studies. The data point to a decreasing mean Eddington ratio with cosmic time and with galaxy stellar mass at fixed redshift. Our data can be reproduced by black holes and galaxies evolving on similar \MBHMS\ relations but progressively decreasing their average Eddington ratios, mean X-ray luminosities, and specific star formation rates, when moving from the starburst to the quiescent phase. Models consistent with the observed \LXMS\ relation and independent measurements of the mean Eddington ratios, are characterised by \MBHMS\ relations lower than those derived from dynamically measured local black holes. Our results point to the \LXMS\ relation as a powerful diagnostic to: 1) probe black hole-galaxy scaling relations and the level of accretion onto black holes; 2) efficiently break the degeneracies between duty cycles and accretion rates in cosmological models of black holes.

\end{abstract}




\section{Introduction}

The origin of the coevolution between central supermassive black holes (BHs) and their host galaxies, most notably mirrored in their scaling relations, persists as an open debate in extra-galactic astronomy. 
The masses of these central BHs
show a correlation with the host galaxy properties, including the stellar mass and velocity dispersion
\citep[e.g.,][]{2013ARA&A..51..511K,2015ApJ...813...82R,2016MNRAS.460.3119S}, 
correlation which is observed to hold even at higher redshifts
\citep[e.g.][]{shankar09c,cisternas11b,Suh20,Li21}.
In particular, the significant interconnection observed between the average X-ray luminosity (L$_{\rm X}$) and the star formation rate (SFR) in active galaxies, i.e. with an active galactic nucleus (AGN), has been often interpreted as a tracer of the elusive underlying link between BH accretion and host galaxy growth across cosmic times \citep[e.g.,][]{2012ApJ...753L..30M,Shankar13Acc}. 
The X-ray luminosity of galaxies is also used as a proxy of black hole 
accretion rate (BHAR) since X-rays are very energetic photons that are created very close to the central BH and
other contaminants in the host galaxies at these wavelengths, for example, emission from stellar processes or binary systems, are usually less powerful and not dominant \citep[e.g.][]{2015A&ARv..23....1B}. 
Many studies have tried to unveil the degree of causality
of the BHAR with host galaxy properties like the SFR and stellar mass (M$_*$) across cosmic time 
by taking advantage of deep surveys and cosmological simulations \citep{2015MNRAS.449..373D, 2015ApJ...800L..10R, 2017ApJ...842...72Y, 2017MNRAS.468.3395M}.
The results point toward a positive correlation between the BHAR and the M$_*$ for star-forming galaxies, 
with a slope close to unity similar to the main sequence (MS) of star forming galaxies and which evolves with redshift.
In particular, \citet{2020A&A...642A..65C} performed a statistical analysis in the COSMOS field in order to study the evolution of the average X-ray luminosity and therefore average BHAR in mass complete samples, as a function of stellar mass, for normal star-forming, quiescent and starburst galaxies up to $z = 3.5$. 
In their work they found that at the highest redshift studied, the three populations of galaxies had similar BHARs, while towards lower redshifts they split among starbursts,
maintaining about constant accretion rates, star forming galaxies, having a decrease in BHAR of $\sim1.5$ orders of magnitude, and quiescent galaxies, characterised by a lower but still significant accretion of their BHs.
It is the aim of the present work to deepen into the observational findings of \citet{2020A&A...642A..65C}, and to provide a physical framework to understand their results in terms of fundamental BH accretion parameters. 

To achieve this goal, we make use of state-of-the-art semi-empirical models (SEMs) of galaxies and BHs in a cosmological context. SEMs are a competitive, fast and flexible methodology, extensively used in recent years to constrain the degree of evolution and mergers in galaxies \citep[e.g.,][]{Grylls20}, as well as the degree of coevolution with their central BHs \citep{2013ApJ...762...70C, 2019MNRAS.487..275G, 2019MNRAS.487.2005C, 2020arXiv201002957A, ShankarNat, Allevato21}. 
The application of SEMs is particularly relevant to the creation of active and normal galaxy ``mock'' catalogues \citep[e.g.][]{2019MNRAS.487..275G, 2019MNRAS.487.2005C, 2020arXiv201002957A, ShankarNat, Allevato21},
which are a vital component of the planning of imminent
extra-galactic surveys such as Euclid \citep[][]{laureijs11}.
SEMs, which by design rely on only a few input assumptions, are particularly effective in providing insightful constraints on the main parameters regulating the dependence of some observables on time or mass.

In this work we perform new estimates of the mean X-ray luminosity \LX\ as a function of galaxy stellar mass, redshift and galaxy life phases, for the X-ray detected sources in the \citet{2020A&A...642A..65C} sample, and use comprehensive semi-empirical mock catalogues of active BHs to pin down which parameters control the shape and evolution of the \LXMS\ relation. We explore a variety of inputs in our model, such as the shape of the Eddington ratio distribution \PLz, which carries information on the accretion of a BH, or the normalisation of the \MBHMS\ scaling relation. We will show in what follows that the slope and normalisation of the \LXMS\ relation are mostly determined by, respectively, the \MBHMS\ relation and the mean Eddington ratio.

In Section~\ref{sec:model} we present our model and in Section~\ref{sec:results} we highlight the main parameters controlling the \LXMS\ evolution at different redshifts and galaxy phases.
In Sections~\ref{sec:disc} and~\ref{sec:concl} we discuss our findings and draw our conclusions on the relevance of our results in the context of the BH-galaxy co-evolution scenario. Throughout this paper we assume a \citet{2003PASP..115..763C} stellar initial mass function, and a flat cosmology with $H_0=70$~Km/s/Mpc, $\Omega_\Lambda=0.7$, $\Omega_M=0.3$.

\section{Building robust AGN mock catalogues}\label{sec:model}
In this study we create realistic mock catalogues of AGN and non-active galaxies to study which input parameters mostly control the \LXMS{} relation at different redshifts. Below we provide the most relevant steps in the generation of our mocks, and refer the reader to \citet{Allevato21} for full details.

The first step for the creation of mocks consists in generating a halo distribution via a halo mass function from \citet{2008ApJ...688..709T} at the redshift of interest. To each dark matter halo we assign a galaxy stellar mass via abundance matching techniques \footnote{We immediately note that the exact choices for this first step of the mock generation are irrelevant to our results and conclusions discussed below. As in this work we are not studying the environment of active galaxies, the information on host halo mass is here only given for completeness. Equivalent mocks could be generated by simply extracting galaxies from an input stellar mass function.}, using the relation of \citet[][]{moster10} with updated parameters from \citet[][Eq.~5]{2019MNRAS.483.2506G} with a normal scatter in stellar mass at fixed halo mass of $0.11$~dex. 
We then assign a BH mass via the empirically calibrated \MBHMS\ relation by \citet{2015ApJ...813...82R}, with an intrinsic scatter of $0.55$ dex, and also explore the impact of adopting other M$_{\rm BH}-{\rm M}_*$ relations from \citet{2016MNRAS.460.3119S}, \citet{2018ApJ...869..113D} and \citet{2019ApJ...876..155S}, which bracket the systematic uncertainties in the BH-galaxy stellar mass in the local Universe. 
We then assume that each relation does not evolve with redshift, as suggested by a number of studies \citep[e.g.][]{shankar09c,2019ApJ...885L..36D, Suh20, Shankar20MNRAS, 2020A&A...642A..65C, Li21, Marsden21}.
To each galaxy and BH we then assign an Eddington ratio $\lambda\equiv L_{bol}/L_{\rm Edd}$ and convert bolometric luminosities $L_{bol}$ to intrinsic (i.e., unobscured) 2-10 keV X-ray luminosities $L_X$ via the same bolometric corrections $k_X$ adopted by \citet{2020A&A...642A..65C}. Following the formalism in, e.g., \citet{Shankar13Acc} and \citet{Allevato21} and references therein, which in turn follows the one routinely adopted in continuity equation models, the AGN luminosity function at any given redshift $z$ can be expressed by the convolution
\begin{equation}
\Phi(\log L_{bol},z)=\int_{\log \lambda_{\rm min}}^{\log \lambda_{\rm max}}U(y,z) n(y,z) P(\log \lambda,z)d\log \lambda \, 
\label{eq:PhiL}
\end{equation}
where $y=\log {\rm M}_{\rm BH}$ and $n(y,z)$ is the total BH mass function. \PLz\ is the Eddington ratio distribution, which we assume for simplicity to be independent of BH mass, normalised to unity in the range $\log \lambda_{\rm min}<\log \lambda<\log \lambda_{\rm max}$. $U(y,z)$ is the \emph{intrinsic} duty cycle, i.e., the fraction of all black holes of mass $y$ that are active and accreting mass at an Eddington rate in the range $\log \lambda_{\rm min}<\log \lambda<\log \lambda_{\rm max}$ at redshift $z$. We set our minimum Eddington ratio to $\log \lambda_{\rm min}=-4$ and the maximum Eddington ratio to $\log \lambda_{\rm max}=1$, noticing that the exact value chosen for $\log \lambda_{\rm max}$ does not alter any of our results as the adopted Eddington ratio distributions have extremely low probabilities above the Eddington limit.

The flexibility offered by Eq.~\ref{eq:PhiL} allows to disentangle the effects of the shape of \PLz, which carries information on the accretion properties of a BH, from the fraction $U(y,z)$ of active BHs accreting above a certain threshold in Eddington ratio. The reference $P(\log \lambda,z)$ distribution is taken to be a simple Gaussian in $\log \lambda$ characterised by a standard deviation $\sigma$ and a mean $\mu$. We will show that the shape of the \PLz\ distribution plays a minor role in the outputs as long as the characteristic Eddington ratio, defined as
\begin{equation}
\zeta_c(z)\equiv\left<\log \lambda\right>(z)=\int_{\log\lambda_{\rm min}}^{\log \lambda_{\rm max}} P(\log\lambda,z)\log(\lambda) \,d\log(\lambda)\, ,
\label{eq:zeta_c}
\end{equation}
is the same. 
We assume a constant duty cycle of $U=0.2$ as suggested by \citet{goulding10} from local X-ray AGN, but we will also explore the impact on our results of varying the input duty cycle with BH mass, specifically decreasing with M$_{\rm BH}$ as inferred by \citet{2010A&A...516A..87S} and \citet{2015MNRAS.447.2085S}, and also increasing with M$_{\rm BH}$, as proposed by \citet{2019MNRAS.488...89M}. Although these works were based on AGN samples with different selections, we use these duty cycles simply as a guidance to explore the impact on our results of different ``shapes'' of the input intrinsic duty cycles $U(y,z)$.

\begin{figure*}
\begin{center}
  \includegraphics[width=0.49\textwidth]{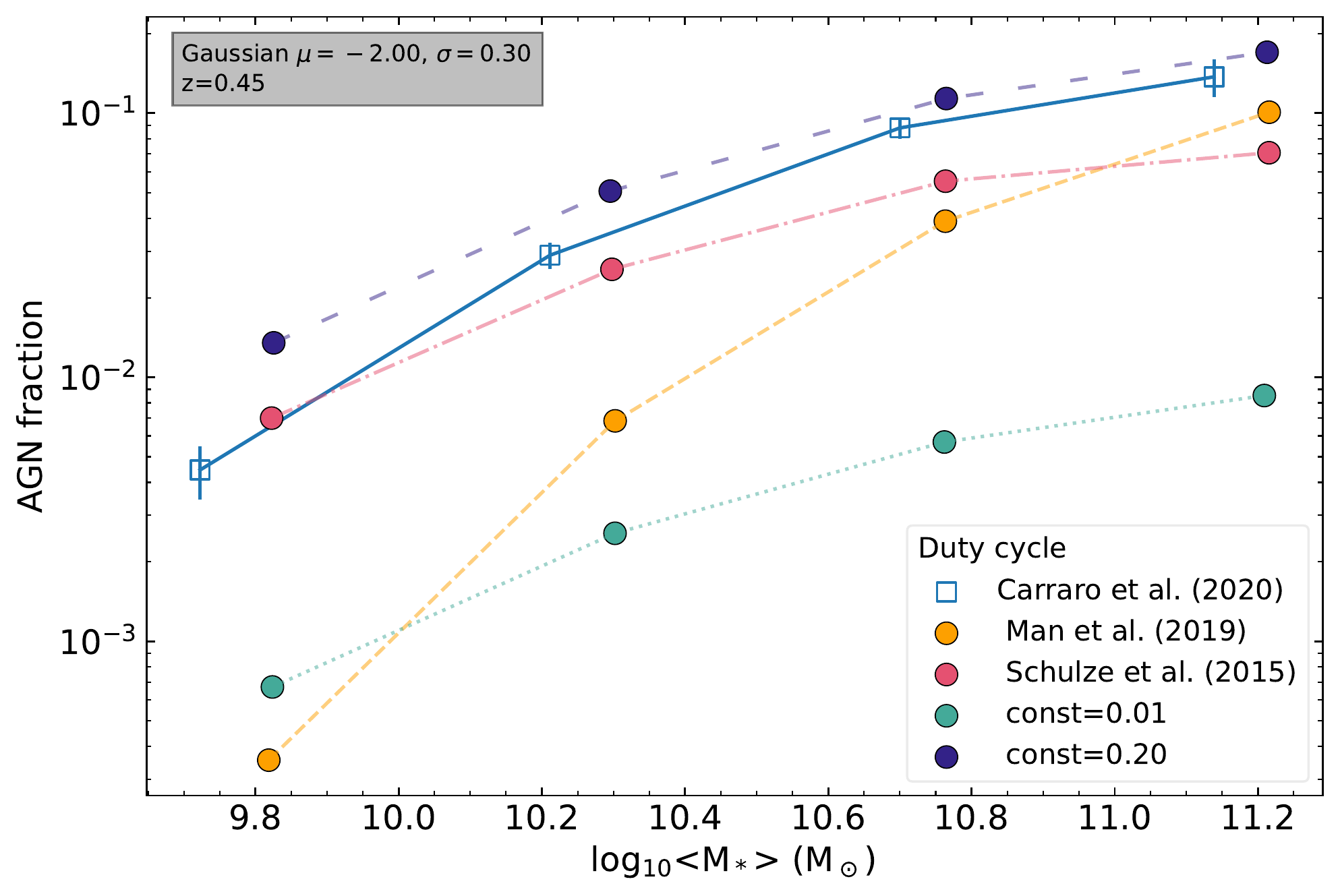}
  \includegraphics[width=0.49\textwidth]{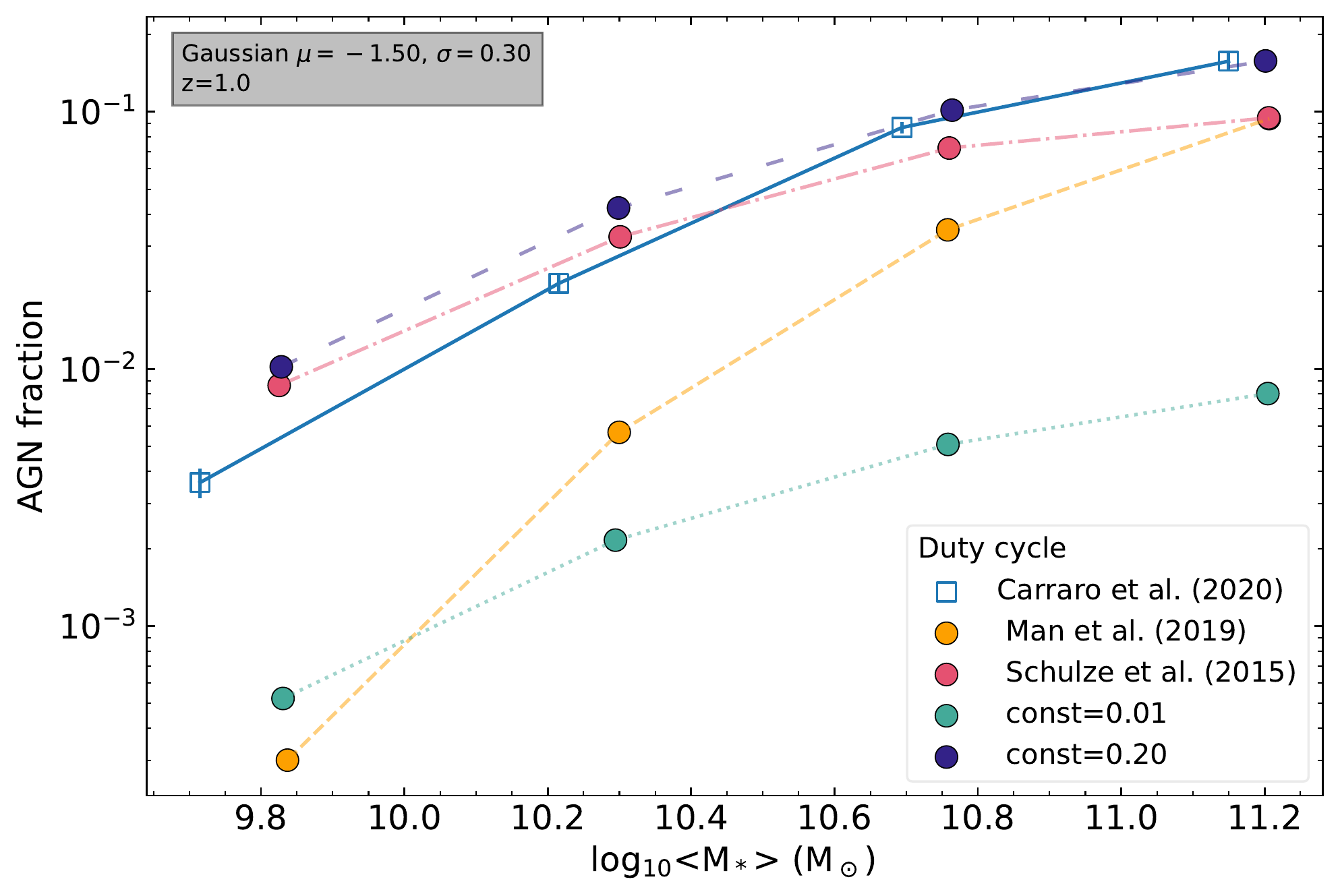}
  \includegraphics[width=0.49\textwidth]{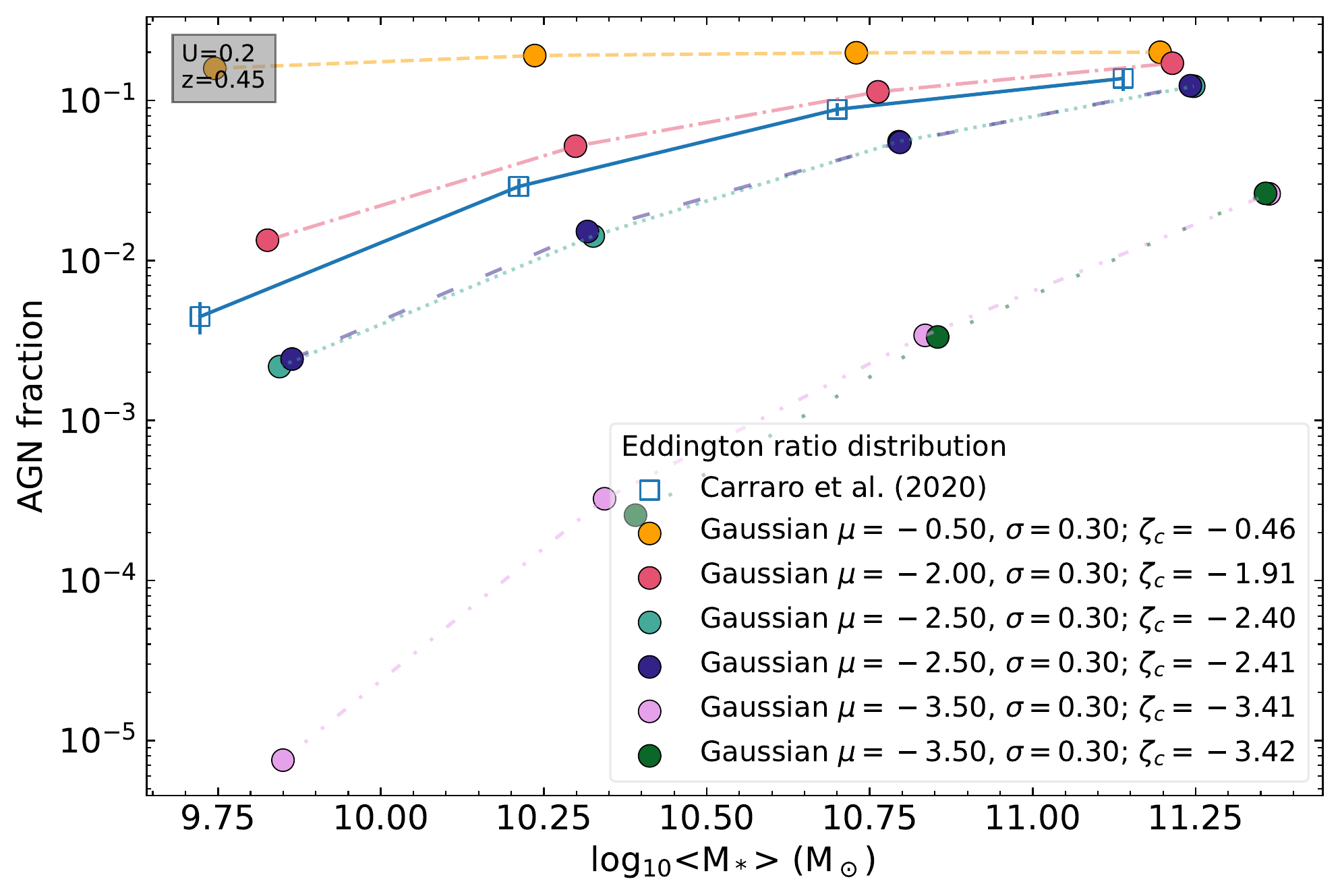}
  \includegraphics[width=0.49\textwidth]{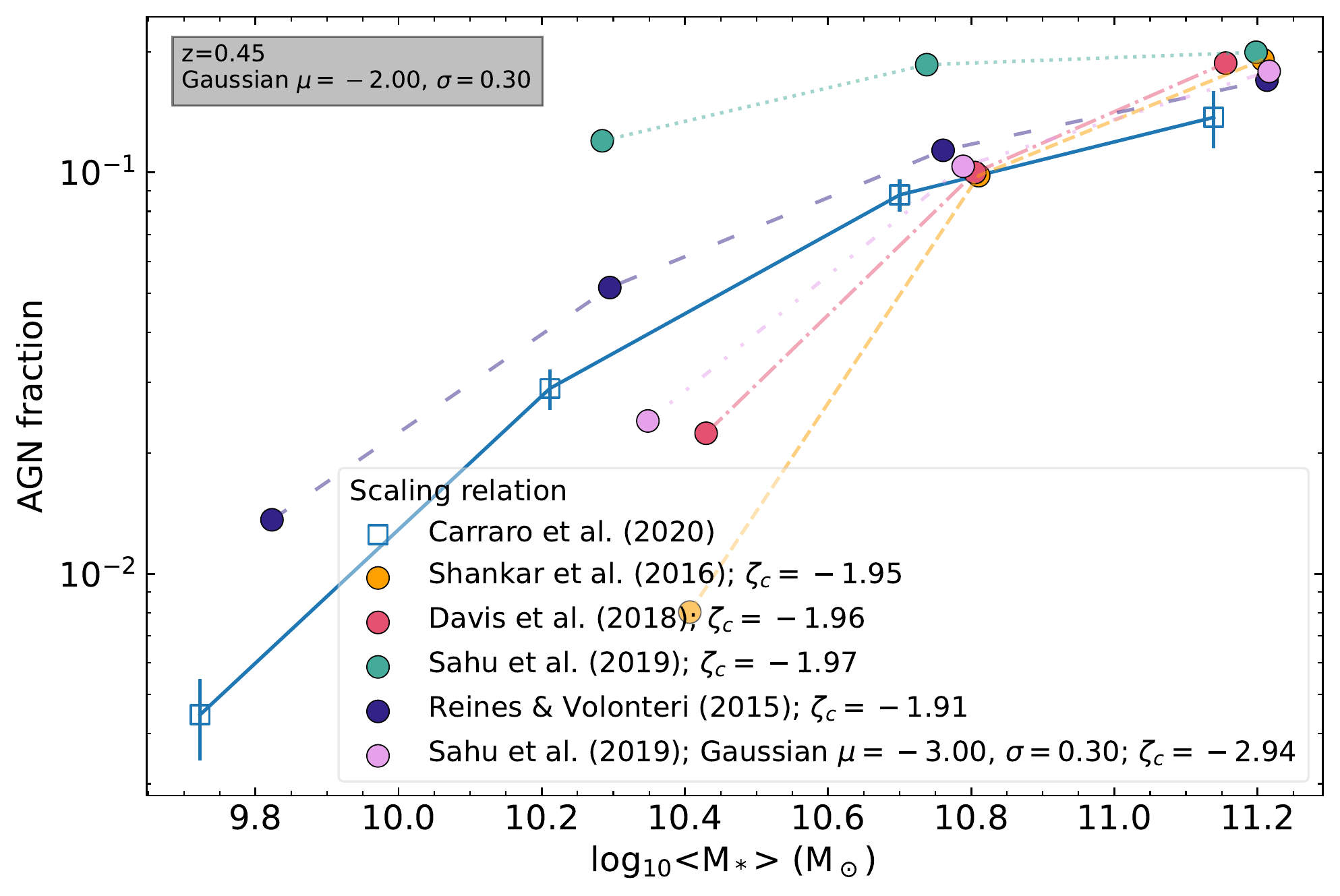}
  \caption{Dependence of the fraction of X-ray detected galaxies (AGN fraction) on the input model duty cycle (top panels),
  Eddington ratio distribution (bottom-left panel) and \MBHMS{} scaling relation (bottom-right panel). We include in each panel the fraction of X-ray detected galaxies from the COSMOS sample from \citet{2020A&A...642A..65C} with an error bar given by the binomial error on the number of detected AGN and a Poisson error on the total number of sources.}
    \label{fig:AGN_fractions}
\end{center}
\end{figure*}

When comparing with the data we must retain from the full BH mock only those active BHs shining above the X-ray flux limit of the observational survey \citep[e.g.,][]{Shankar13Acc}. In our reference sample, the \textit{Chandra} COSMOS Legacy Survey \citep[COSMOS-Legacy,][]{2016ApJ...819...62C}, the X-ray flux limit corresponds to luminosities of $L_X=10^{42}\, {\rm erg/s}$ in the lower redshift bin ($z=0.45$), increasing by an order of magnitude or more at higher redshifts (see below for details). When computing all AGN-related observational probes, such as the AGN luminosity function (Eq.~\ref{eq:PhiL}), the characteristic Eddington ratio $\zeta_c(z)$ (Eq.~\ref{eq:zeta_c}), or the mean X-ray luminosity (Eq.~\ref{eq:meanLx}), we thus include only those active black holes shining above the \textit{Chandra} COSMOS Legacy Survey flux limit at the given redshift{\footnote{The COSMOS field is a great combination of area and depth making it ideally suited to probe the accretion properties of active BHs. A deeper field may be more sensitive to the faint end shape of the Eddington ratio distribution, but would not allow to include more luminous sources. A shallower field on the other hand, may return better statistics for the more luminous sources, but rapidly losing the fainter ones.}}. For example, although we fix our minimum Eddington ratio to $\log \lambda_{\rm min}=-4$ for our input \PLz\ (e.g., in Eq.~\ref{eq:PhiL}), after imposing the cut in X-ray flux limit, among the BHs with mass ${\rm M}_{\rm BH} \lesssim 10^8\, M_{\odot}$ in the lowest redshift bin, only those accreting at an Eddington rate $\log \lambda_{\rm X, min} \gtrsim -3$ will be included in the comparison with the data. We will discuss below that the flux limit plays a non-negligible role when comparing theoretical AGN mocks to observations, particularly with respect to the observed fraction of active black holes as a function of host galaxy stellar mass (Figure~\ref{fig:AGN_fractions}).

We assign SFRs to quiescent, normal star-forming, and starburst galaxies based on their respective SFR-M$_*$ relation.
For starburst and quiescent galaxies, we adopt the SFR fits from
\citet[Table~3]{2020A&A...642A..65C}, while for the ``main sequence'' we adopt the 
\citet[Eq.~9]{2015A&A...575A..74S} flexible parametric formula
\begin{equation}
    \log_{10}\left(\frac{\rm SFR}{M_\odot yr^{-1}}\right)= m - m_0 +a_0 r - a_1[\max(0,m-m_1-a_2r)]^2
	\label{eq:SFR}
\end{equation}
with $m\equiv\log_{10}(M_*)-9$ and $r\equiv\log_{10}(z+1)$. Best-fit parameters for our COSMOS data are $a_0=2.29\pm 0.12$, $a_1=0.25 \pm 0.04$, $a_2=0.33 \pm 0.30$, $m_0=0.64 \pm 0.03$, $m_1=0.55\pm 0.11$. We add a dispersion of $0.2$~dex to the SFR. 

Irrespective of their duty cycle, we assign to each galaxy in the mock an X-ray luminosity from X-ray binary emission following \citet[][Table 3]{2016ApJ...825....7L}, and when computing the average X-ray luminosity competing to a given bin of stellar mass, we then subtract the mean binary emission competing to that bin of stellar mass and star formation rate, strictly following the same methodology pursued in \citet{2020A&A...642A..65C}. We note that neglecting X-ray binary emission entirely from our procedure would yield very similar results.
Following the procedure described above, we generate diverse galaxy mock catalogues with distinct choices of the input \MBHMS{} scaling relations, duty cycles, and $P(\log\lambda,z)$ distributions. 
We then divide each AGN mock catalogue in bins of stellar mass, and select the BHs that shine above the flux limit of the COSMOS-Legacy survey \citep{2016ApJ...817...34M}, i.e., the ``detected'' sources of the mock, as discussed above. The first observable we compute is the 
AGN fraction, defined as
\begin{equation}
{\rm AGN\;fraction}(M_*,z)=\frac{\sum_i U_{i,{\rm detected}}} {N_{\rm tot}}\, ,
    \label{eq:AGN_fraction}
\end{equation}
where the sum in the numerator runs over all active BHs above the flux limit, and $N_{\rm tot}$ at the denominator is the total number of active and normal galaxies in the specified stellar mass bin. We note that the probability for a galaxy to be detected above a certain X-ray luminosity threshold, i.e., the ``observed'' duty cycle, will depend not only on the assumed (intrinsic) duty cycle, but also on other properties such as its BH mass and Eddington ratio. We will discuss below the differences between observed and intrinsic duty cycles and highlight how different input parameters in the mocks can generate similar observed fractions of AGN.
The comparison between the observed AGN fraction and predicted input duty cycle \UMBHz\ yields important constraints on the accretion properties of active BHs when coupled to other observables, as we will discuss below.
Finally, we perform 500 bootstraps out of which we extract the median SFR and M$_*$, and the linear mean L$_{\rm X}$ weighted by the AGN duty cycle \begin{equation}
\left<L_X\right>=\frac{\sum_i U_i(y_i,z) L_X(y_i)} {\sum_i U_i(y_i,z)}\, ,
    \label{eq:meanLx}
\end{equation}
where $\log L_X(y_i)=38.1 +\log \lambda_i +y_i-\log k_X$, where again the sums run over all detected BHs in the selected stellar mass bin. The key advantage of computing mean X-ray luminosities only considering sources above the flux limit, is that it provides a tracer of BH luminosity largely independent of the duty cycle, as demonstrated below. For each bootstrapped distribution we compute the median SFR and \LX\ with their 5th and 95th percentiles, following the same procedure as in the comparison observational sample selected by \citet{2020A&A...642A..65C}. We note that in the original work by \citet{2020A&A...642A..65C} the mean X-ray luminosities were computed over the full sample of ${\rm M}_*$-selected galaxies, including both detected sources and stacking on non-detected sources. Eq.~\ref{eq:meanLx} is instead a weighted mean over only the detected sources, and thus we recomputed the mean X-ray luminosities in the \citet{2020A&A...642A..65C} sample limiting the analysis to only X-ray detected sources. Whilst \citet{2020A&A...642A..65C} assumed an average characteristic obscuration/extinction correction for all sources competing to a given bin of X-ray luminosity, we here apply to each individual source the obscuration correction listed in the \citet{2016ApJ...817...34M} catalogue.
These new estimates of the mean X-ray luminosities, which will be presented below, will be used in what follows as term of comparison for our models. 
We checked that the weighted X-ray luminosity in a given bin of stellar mass given by Eq.~\ref{eq:meanLx} is equivalent to the simple arithmetic mean $\left<L_X\right>=\sum_i L_X(y_i)/N_{\rm AGN}$ over a randomly selected subsample of galaxies $N_{\rm AGN}=U\times N_{\rm tot}$ in the given bin of stellar mass (in fact, the two expressions become formally identical in the limit of strictly constant duty cycles). The advantage of adopting a weighted mean over a simple arithmetic one is that the former is more stable against numerical noise induced by low number statistics. A detailed comparison to the data would require to distinguish in the model the relative fractions of starburst, main sequence and quiescent galaxies, and for each galaxy class compute a distinct mean X-ray luminosity via Eq.~\ref{eq:meanLx}, with the sum running over the subsample of detected sources competing to the specified galaxy type. However, such fractions would appear as additional constant weights in both the numerator and denominator of Eq.~\ref{eq:meanLx}, and as such they would be cancelled out. In other words, we checked that computing a mean $\left<L_X\right>=\sum_i L_X(y_i)/N_{\rm AGN}$ over all detected sources irrespective of their  star-forming type, or restricting the calculation of the mean to only the relative numbers of detected sources per galaxy type as observed in the data \citep[][their Tables A.1, A.2, A.3]{2020A&A...642A..65C}, yields equivalent results for the same set of input parameters.

\section{Results}\label{sec:results}

\begin{figure*}
\begin{center}
  \includegraphics[width=\textwidth]{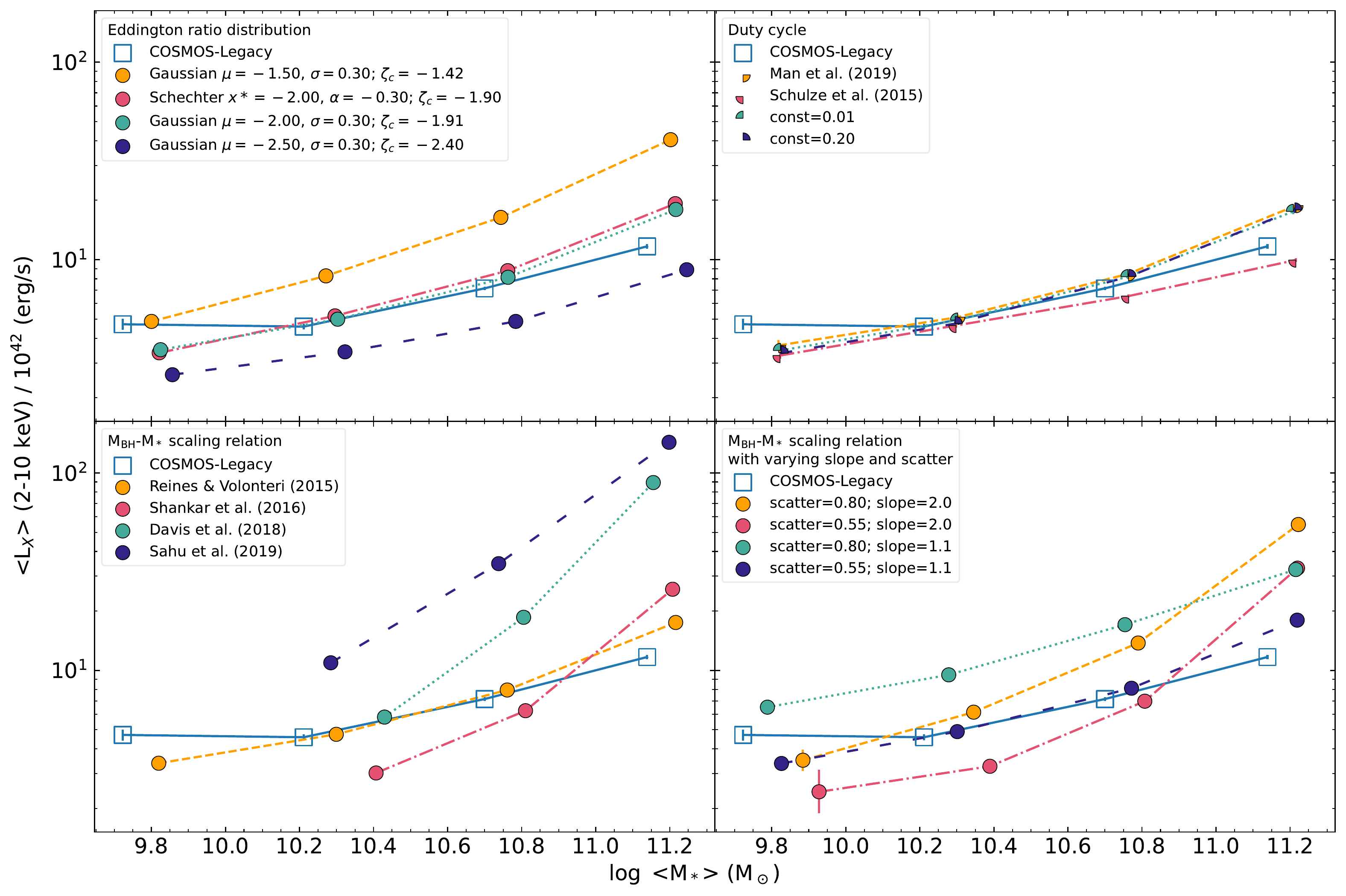}
  \caption{A gallery of average  L$_{\rm X}-{\rm M}_*$ relations of detected sources at $z=0.45$ for star forming galaxies obtained by varying one of the input relations at a time. The relation that varies in each subplot is reported in the legend. 
  Results from COSMOS X-ray detected sources at the same redshift from \citet{2020A&A...642A..65C} are included in all plots for comparison (blue squares).
  Top left: L$_{\rm X}-{\rm M}_*$ relation obtained by changing the Eddington ratio distribution function. We use a Schechter function and Gaussian function in $\log(\lambda)$ with varying mean $\mu$ and standard deviation $\sigma$ values.
  Top right: L$_{\rm X}-{\rm M}_*$ relation obtained by changing the duty cycle method.
  Bottom left: L$_{\rm X}-{\rm M}_*$ relation obtained by changing the M$_{\rm BH}-{\rm M}_*$ scaling relation. Each scaling relation is shown within its original stellar mass range of derivation.
  Bottom right: L$_{\rm X}-{\rm M}_*$ relation obtained with a toy M$_{\rm BH}-{\rm M}_*$ scaling relation where we change the logarithmic slope $\beta$ of the relation $\log {\rm M}_{\rm BH} = \alpha + \beta \log {\rm M}_*$ and increase its scatter. Original \citet{2015ApJ...813...82R} values are: $\beta=1.1$ and 0.55 dex scatter.
  }
    \label{fig:LX_M}
\end{center}
\end{figure*}
\subsection{Reproducing the measured fraction of detected galaxies} \label{ssec:Fig1}

Before showing the results on the predicted mean X-ray luminosity of detected galaxies, we discuss if and when our model is able to match the fraction of X-ray sources directly observed in COSMOS-Legacy as a function of stellar mass. The open blue squares in Figure~\ref{fig:AGN_fractions} are the COSMOS Legacy data, taken from Table A.1 in \citet{2020A&A...642A..65C} to which we associate a binomial error on the number of detected AGN and a Poisson error on the total number of sources, combined together with standard error propagation applied to $\frac{N_{\rm det}}{N_{\rm tot}}$. 
We then compare the data with our models filtered by the flux limit of the observations, which is equal to ${\rm L}_{X,min}=10^{42}erg/s$ and ${\rm L}_{X,min}=6\times10^{42}erg/s$ at $z=0.45$ and $z=1.0$, respectively. We adopt as our reference model one characterised by a constant input duty cycle of $U=0.2$, a Gaussian Eddington ratio distribution in $\log\lambda$ peaked at $\mu=-2$, and the \MBHMS{} scaling relation from \cite{2015ApJ...813...82R}. 
We will show below that this choice of input parameters provides a good match to both the mean AGN X-ray luminosity and AGN luminosity function. We then vary several of the input parameters, starting from the duty cycle at both $z=0.45$ and $z=1$ (left and right top panels, respectively), the peak of the Gaussian \PLz\ (bottom, left panel), and the input \MBHMS{} scaling relation (bottom, right panel). It is first of all interesting to note from the top panels that, once the Gaussian $P(\log \lambda,z)$ and \MBHMS\ scaling relation are fixed to our reference choices, the data are consistent with an input duty cycle $U\sim0.2$ constant in both stellar/black hole mass and redshift, at least up to $z\lesssim 1$ (dark blue dashed lines in both top panels). The apparent strong increase of the AGN fraction with stellar mass is simply induced by the imposed flux limit. A too strong mass dependence in the input duty cycle, as suggested by the local fraction of optical AGN measured by \citet{2019MNRAS.488...89M} in SDSS, would be inconsistent with the data (dashed, orange lines), as well as an overall too low initial fraction (dotted, turquoise lines with $U=0.01$). 

The bottom left panel of Figure~\ref{fig:AGN_fractions} shows that a varying input \PLz{} distribution, and thus a varying characteristic $\zeta_c$, as labelled, generates widely different AGN fractions. More specifically, the higher the $\zeta_c$ the more luminous are, on average, the mock AGN, which in turn implies that proportionally less sources are removed by the cut imposed by the flux limit. We find that when $\zeta_c \gtrsim -0.5$, the observed AGN fraction is nearly identical to the input $U\sim 0.2$ (dashed, yellow line), while it rapidly {diverges from the input $U\sim 0.2$ dropping towards lower mass, less luminous AGN when $\zeta_c \lesssim -2$}.
The right lower panel of Figure~\ref{fig:AGN_fractions} also shows that a flatter or steeper \MBHMS\ input scaling relation, such as the ones from dynamically measured $M_{\rm BH}$ by \citet[][dotted, turquoise line]{2019ApJ...876..155S} in early type galaxies and \citet[][dot-dashed, magenta line]{2018ApJ...869..113D} in late type galaxies, naturally induce a proportionally flatter or steeper AGN fraction, because they map galaxies of same stellar mass to more massive/more luminous or less massive/less luminous AGN. In conclusion, the observed AGN fraction can contribute to efficiently break the degeneracies in the input parameters (see also Section~\ref{sec:disc}), and, when combined with other independent constraints on, e.g., the BH-galaxy scaling relations and/or the Eddington ratio distributions, it is a powerful diagnostic of the intrinsic AGN duty cycle $U(y,z)$, and it can thus be used to constrain the accretion history of supermassive black holes. 

\subsection{The effect of the model's inputs on the \LXMS{} relation} \label{ssec:Fig2}

In Figure~\ref{fig:LX_M} we compare the mean X-ray luminosity of detected active galaxies in a given bin of stellar mass, which in what follows we will continue labelling simply as \LX{} (Eq~\ref{eq:meanLx}), with several different model predictions. To pin down the input parameters that mostly control the \LXMS\ relation, we explore in Figure~\ref{fig:LX_M} how the relation varies by changing, from top left to bottom right, the \PLz\ , the duty cycle, the full \MBHMS\ relation, and only the slope/scatter of the \citet{2015ApJ...813...82R} relation, as labelled. All the mocks are generated at $z=0.45$, though the results are applicable to all redshifts, as further discussed below. In Figure~\ref{fig:LX_M} the data refer to only the subsample of star forming, main sequence galaxies. As anticipated in Sec~\ref{sec:model} and Eq~\ref{eq:meanLx}, the mean \LX\ should in principle be weighted by the fractional number of detected sources within a given star formation class (e.g., quiescent, star forming, starbursts). However, this additional weighting can be neglected as it is cancelled out in Eq~\ref{eq:meanLx}, being a constant in each bin of stellar mass \citep[e.g.,][]{2020A&A...642A..65C}.

\begin{figure*}
\begin{center}
  \includegraphics[width=\textwidth]{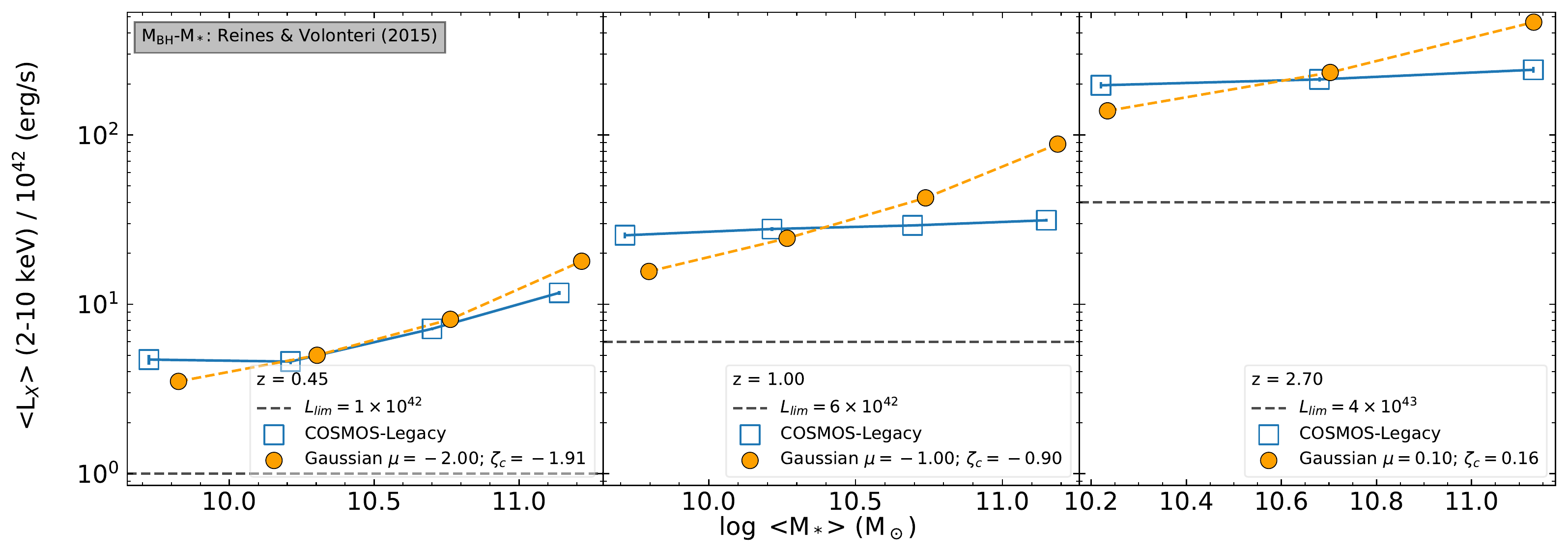}
  \caption{The L$_{\rm X}-{\rm M}_*$ relations at $z=0.45$ (left panels), $z=1.0$ (central panels) and $z=2.7$ (right panels) obtained by assuming a M$_{\rm BH}-{\rm M}_*$ scaling relation from \citet{2015ApJ...813...82R} and a Gaussian in $\log (\lambda)$ with standard deviation $\sigma=0.3$~dex. We vary the Eddington ratio distribution in order to reproduce the observational results from the COSMOS-Legacy detected sources selected in \citet{2020A&A...642A..65C}. Black dashed lines represent the survey luminosity limits. }

    \label{fig:LX_M_redshift}
\end{center}
\end{figure*}

The top left panel of Figure~\ref{fig:LX_M} compares the mean \LX\ measured in the data with the one from our mocks of detected galaxies. We find that a $\zeta_c \sim -2$ is able to match the data at $z=0.45$. This value of the mean Eddington ratio is broadly consistent with the mean specific BH accretion rate $\lambda_{\rm sBHAR}$ measured by \citet{2019MNRAS.484.4360A} from large samples of deep X-ray AGN surveys, and also with the mean Eddington ratio quoted by other groups \citep[e.g.,][]{H09,Kauff09}. The top panels of Figure~\ref{fig:LX_M} clearly show that whilst the normalisation of the \LXMS\ relation is strongly controlled by the characteristic Eddington ratio $\zeta_c$ (left panel), it has a negligible dependence on the input AGN duty cycle (right panel). This behaviour is expected as the \LX\ in Eq~\ref{eq:meanLx} is an average luminosity calculated only on the fraction of detected sources, and as such it is largely independent of the number of BHs detected in a given bin of stellar mass, but strongly dependent on the {\emph{rate}} at which these BHs are accreting. We show in the top left panel of Figure~\ref{fig:LX_M} that a Schechter or Gaussian \PLz\ yield the same mean X-ray luminosity \LX\ at fixed stellar mass as long as their $\zeta_c$ are the same (dotted turquoise and dot-dashed magenta lines). It is indeed the characteristic Eddington ratio $\zeta_c$, and not the overall shape of the \PLz\ input distribution, to determine the level of mean X-ray luminosity in detected galaxies at fixed stellar mass and
at fixed \MBHMS\ relation. Nevertheless, some constraints even on the shape of the \PLz\ can be derived from our methodology. For example, assuming a steeper/flatter faint end in the input Schechter \PLz\ function, would induce a lower/higher $\zeta_c$. To then preserve the same $\zeta_c$ necessary to match the observed \LXMS\ relation, would in turn require a shift in the knee of the Schechter function, and the new combination of faint end slope and knee can then be tested against the AGN luminosity function (which we further discuss below). It is relevant to reiterate at this point that the observations are only sensitive to Eddington ratios corresponding to luminosities above the survey flux limit (Eq.~\ref{eq:zeta_c}), and thus are sensitive only to portion of the \PLz\ above the minimum Eddington ratio detectable in the sample.

Some residual, weak dependence on the duty cycle may be visible in the right panel of Figure~\ref{fig:LX_M} especially towards higher stellar masses (dote-dashed, magenta line). This (tiny) dependence of the \LXMS\ relation on the input duty cycle is a simple byproduct of the scatter in the \MBHMS\ relation and of our definition of input duty cycle: \UMBHz\ is dependent on BH mass, and thus at fixed stellar mass, a variety of BHs with different weights could contribute to the mean \LX{}, slightly altering its final value depending on the shape (not the normalisation) of the input duty cycle \UMBHz{}.

The bottom left panel of Figure~\ref{fig:LX_M} shows instead a close link between the normalisation of the input \MBHMS\ relation and the normalisation in the \LXMS\ relation: at fixed $\zeta_c$, a lower \MBHMS\ relation will result in a proportionally lower \LXMS\ relation, and vice versa. This link between the two relations naturally arises from the proportionality between X-ray luminosity and BH mass, which in turn is linked to the host galaxy stellar mass via the \MBHMS\ relation. The right panel of Figure~\ref{fig:LX_M} shows the variations in the \LXMS\ relation for the same input \MBHMS\ relation with varying slope or scatter, as labelled. A steeper/shallower \MBHMS\ scaling relation will result in a proportionally steeper/shallower \LXMS\ relation, while a lower/higher scatter will decrease/increase the normalisation of the \LXMS\ relation, mainly due to the lower/larger contribution of detected BHs, especially the more massive and luminous ones. It is thus clear from Figure~\ref{fig:LX_M} that the slope and normalisation of the input \MBHMS\ relation, as well as the input $\zeta_c$, all play a significant, and in fact degenerate, role in shaping the \LXMS\ relation. For example, a flatter slope in the \MBHMS\ relation or a mass-dependent $\zeta_c$, progressively decreasing at larger masses, could both produce a flatter slope in the \LXMS\ relation. Also, decreasing $\zeta_c$ with increasing BH mass could indeed reconcile the \citet{2020A&A...642A..65C} observational results with a steeper \MBHMS\ relation as calibrated in the local Universe \citep[e.g.,][]{2016MNRAS.460.3119S,2018ApJ...869..113D}. If the scaling relation between BHs and their hosts is constrained via independent methods, such as AGN clustering \citep[e.g.,][]{ShankarNat,Allevato21,Viita21}, then the \LXMS\ relation can be used to constrain the mean $\zeta_c$ as a function of galaxy stellar mass and redshift, as further discussed below.

\begin{figure*}
\begin{center}
\includegraphics[width=\textwidth]{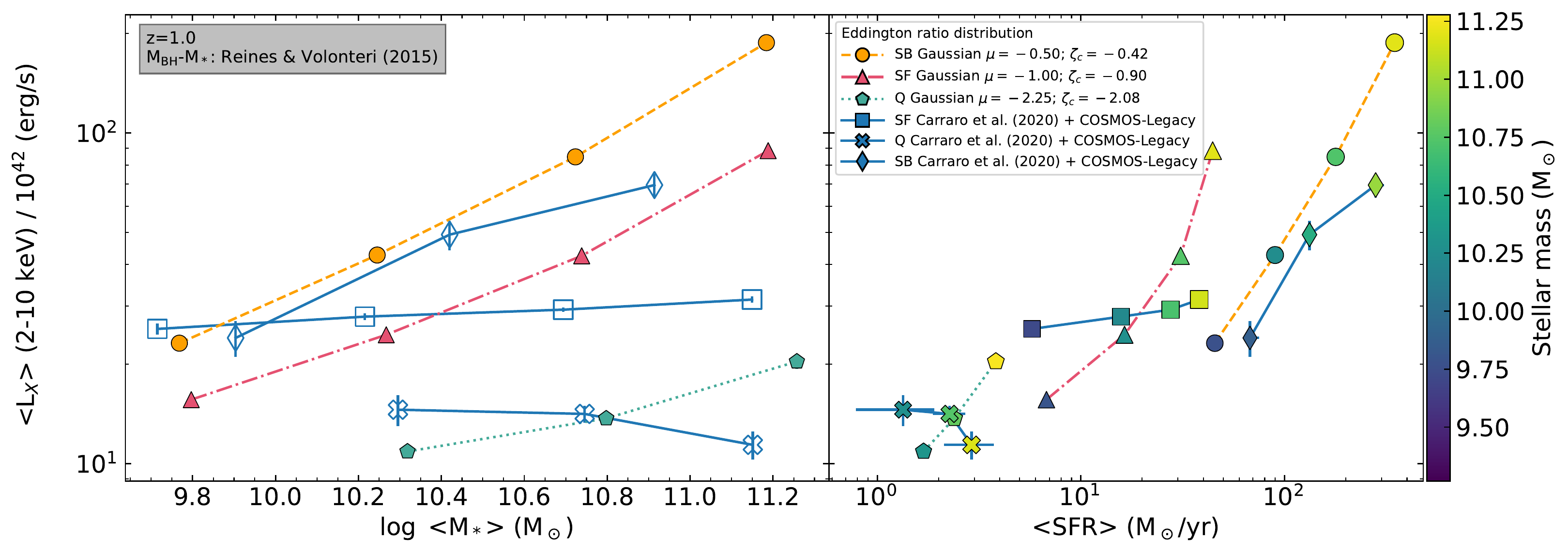} 
  \caption{L$_{\rm X}$ as a function of ${\rm M}_*$ (left) and SFR (right). L$_{\rm X}$ are obtained at 
  $z=1.0$ for detected galaxies with a \citet{2015ApJ...813...82R} M$_{\rm BH}-{\rm M}_*$ scaling relation and with a Gaussian Eddington ratio distribution as shown in the legend, with a $\sigma=0.3 dex$. In the right panel, SFRs are obtained using the fits from \citet{2020A&A...642A..65C} for star-forming (SF), quiescent (Q) and starburst (SB) galaxies, and data points are colour coded according to ${\rm M}_*$. All relations are compared with results from COSMOS data from \citet{2020A&A...642A..65C}.}
    \label{fig:SFQSB_active}
\end{center}
\end{figure*}

\subsection{Reproducing the \LXMS\ relation through cosmic time}

In this Section we extend the comparison to data on the \LXMS\ relation at different redshifts. We showed in Section~\ref{ssec:Fig2} that the \LX\ can provide valuable constraints on the mean Eddington ratio of active BHs. Thus, by studying the \LXMS\ at different redshifts and galaxy stellar masses, we can build a more comprehensive view of how BHs accrete at different epochs and in different host galaxies.
The data point to a steady decrease of the mean \LXMS\ luminosity with cosmic time at fixed host galaxy stellar mass. As discussed above, this decreasing trend could be interpreted either as a progressive decline in the normalisation of the \MBHMS\ relation and/or in the characteristic $\zeta_c$. The latest data suggest a rather weak evolution in the \MBHMS\ relation up to at least $z\sim 2.5$ \citep[e.g.,][]{Suh20,Shankar20MNRAS} thus favouring, in our approach, a steady decrease in $\zeta_c$, which would also be in line with independent observations \citep[e.g.,][]{Kollmeier06} and continuity equation models \citep[e.g.,][]{Shankar13Acc,Aversa15}.

In Figure~\ref{fig:LX_M_redshift} we show the predicted \LXMS\ relation for mock catalogues at $z=0.45,1.0,2.7$ (left, central and right panels respectively), generated by assuming as a reference the \citet{2015ApJ...813...82R} \MBHMS\ relation, which naturally generates a slope in the \LXMS\ relation consistent with our data. 
At each redshift we plot the models with an input Gaussian distribution \PLz\ with a $\mu$ value (the corresponding $\zeta_c$ values are very similar being Gaussian distributions) chosen in a way to match the central value of the \LXMS\ distribution at each redshift.
We find that, assuming a strictly constant \MBHMS\ relation, to reproduce the data we would need a drop of a factor $\gtrsim 100$ in the characteristic Eddington ratio $\zeta_c$ from $z\sim 2.7$ to $z\sim 0.45$, which mirrors the fast drop in mean Eddington ratio also derived in some observational data and continuity equation results \citep[see, e.g., Fig. 12 in][]{Shankar13Acc}. We checked that the steady decrease in $\zeta_c$/$\mu$ with decreasing redshift is not an artefact of the progressively lower flux limit with cosmic time (dashed, horizontal lines in Figure~\ref{fig:LX_M_redshift}). Recomputing $\zeta_c[z]$ imposing the same flux limit in all redshift bins yields very similar results. We also note that at $z \gtrsim 1$, on the assumption that the input \MBHMS\ relation remains constant in both slope and normalisation, the models tend to produce a \LXMS\ relation steeper than what observed, which in turn would require a $\zeta_c$ decreasing with increasing stellar mass by a factor $\lesssim 3$ to improve the match to the data. A systematically lower mean Eddington ratio for more massive galaxies would imply that their more massive BHs should have grown earlier, the so-called {\emph downsizing} trend, in which more massive galaxies/BHs build up the bulk of their mass faster than less massive galaxies/BHs \citep[e.g.,][]{Marconi04}. It is important to highlight that the amount of downsizing/decrease in $\zeta_c$ with increasing host galaxy stellar mass, would be reduced if one includes in the mocks a larger statistical uncertainty in stellar mass and/or X-ray luminosity, which would both tend to flatten the predicted \LXMS\ relation \citep[see, e.g., discussion in][]{Shankar14,Allevato19}. All in all, the results in Figure~\ref{fig:LX_M_redshift}, taken at face value, suggest that BHs would be accreting close to their Eddington limit at $z\gtrsim 2.5$, and then rapidly shut off at lower redshifts, especially for more massive galaxies. Indeed, continuity equation models clearly show that more massive BHs have formed most of their mass by $z\sim 1$ \citep[e.g.,][]{Marconi04,Shankar20MNRAS}.  

\subsection{Reproducing the \LXMS\ relation in starburst, main-sequence and quiescent galaxies} \label{subsec:SFQSB}

So far we have mostly focused on comparing model predictions with the mean \LXMS\ relation of star forming main sequence galaxies, which are the vast majority of the detected active galaxies in COSMOS-Legacy. However, AGN activity is routinely detected also in other galaxy life phases. \citet{2020A&A...642A..65C} showed that, at least at $z<2.25$, starbursts, star forming and quiescent galaxies are characterised by distinct \LXMS\ relations, which are similar in slope but differ in normalisation by a factor of $\sim 10$ when moving from quiescent galaxies, with the lowest average \LX, to the starbursts, with the highest average \LX\ at fixed stellar mass. In the context of our approach, this offset in \LX\ at fixed stellar mass could be explained either by a systematic difference in the characteristic Eddington ratio $\zeta_c$ and/or by a systematic offset in the normalisation of the \MBHMS\ relation, when moving from quiescent to star-forming galaxies. In this Section we proceed with a systematic comparison of our models with the COSMOS-Legacy data focusing on the dependence of the \LXMS\ relation on galaxy type at fixed redshift, specifically at $z=1$, though the conclusions we will retrieve below are quite general and can be easily extended to other redshift bins.
 
In the left panel of Figure~\ref{fig:SFQSB_active} we explore mocks with a constant input \MBHMS\ relation from \citet{2015ApJ...813...82R}, but characterised by distinct $\zeta_c$, as labelled (circles, triangles, and pentagons), against the \LXMS\ relation measured for the three types of galaxies studied by \citet{2020A&A...642A..65C} (blue diamonds, squares and crosses for starbursts, star forming, and quiescent galaxies, respectively). Reproducing the steep increase in mean \LX\ at fixed ${\rm M}_*$ requires, as expected, a proportionally higher value of $\zeta_c$ in main sequence and starburst galaxies, assuming the same \MBHMS\ relation. We stress that the calculation of the mean \LX\ of each galaxy type via Eq.~\ref{eq:meanLx} would require an additional statistical weight specifying the relative contribution of each galaxy type to the total number of detected active galaxies. As this weight is constant in each stellar mass bin, it would however cancel out when applied to the numerator and denominator of Eq.~\ref{eq:meanLx}.
In the right panel of Fig.~\ref{fig:SFQSB_active} we show the SFRs of the entire sample from \citet{2020A&A...642A..65C} against the luminosity of the X-ray detected sources only. We decided not to use the SFRs from the detected sample since the tracers used for their estimation (IR and UV luminosity) may be polluted by AGN emission and the IR stacking may not achieve a significant signal-to-noise with the low number statistics from this subsample, both leading to non-representative SFRs for these galaxies.

Interestingly, it is apparent from Figure~\ref{fig:SFQSB_active} that the observed \LXMS\ relation in starburst galaxies is not a simple power law but tends to show a break that becomes more pronounced in more massive galaxies of mass $\log (M_*/M_{\odot}) \gtrsim 10.5$ and at lower redshifts. In our modelling, this feature could be naturally reproduced with a further decrease in $\zeta_c$ in the most massive galaxies in our sample, which would align with the idea of downsizing, as discussed above. This result supports the view that, already in the early starburst phase, more massive galaxies and their central BHs have accreted their mass earlier and are now in their declining phase, as predicted by some models \citep[e.g.,][]{Lapi18}. We stress that the downsizing in $\zeta_c$ would be even more pronounced if steeper \MBHMS\ relations were adopted in input. The right panel of Figure~\ref{fig:SFQSB_active} shows that our chosen values of $\zeta_c$ that match the \LXMS\ relation for each galaxy type, also reproduce, at the same time, their respective ${\rm L}_{\rm X}-{\rm SFR}$ relations, where the SFR is assigned to each galaxy type based on their observed underlying ${\rm SFR}-{\rm M}_*$ relation. 

An alternative way to explain the different normalisation of starburst and quiescent galaxies in the \LXMS\ plane would be to adopt the same $\zeta_c$ for all galaxy types and progressively increase the normalisation of the \MBHMS\ scaling relation when moving from quiescent to starburst galaxies. We however disfavour such a model. Direct measurements of the \MBHMS\ scaling relation in AGN within a variety of host galaxies \citep[e.g.,][and references therein]{2015ApJ...813...82R,Shankar19,Suh20}, have all revealed normalizations that are lower than those typically measured locally in dynamically measured BHs, possibly due to some biases in the latter \citep[e.g.,][]{2016MNRAS.460.3119S}. In particular, BHs in local elliptical, quiescent galaxies seem to be the most massive ones at fixed host galaxy stellar mass among all samples of local active and normal galaxies (see, for example, Figure 8 in \citealt{2015ApJ...813...82R}). In addition, also the analysis of the clustering of active, mostly star-forming, galaxies at fixed BH mass favours \MBHMS\ scaling relations with a normalisation lower than the one measured for local quiescent, early-type galaxies \citep[e.g.,][]{ShankarNat,Allevato21,Viita21}. Direct (or indirect) comprehensive measurements of the \MBHMS\ scaling relation in active starburst galaxies are still unavailable. However, theoretical models suggest that the ratio between BH mass and host galaxy stellar mass in the starburst phase should, if anything, be lower than what observed locally, as the BH grows from a relatively small seed within a host forming stars at exceptionally high rates \citep[see, e.g.,][their Figure 3]{Lapi14}. More generally, these models suggest that, from an evolutionary point of view, quiescent galaxies should be older galaxies with larger BHs at fixed stellar mass \citep[e.g.,][]{Ciras05,Granato06,Lapi06,shankar06,Lapi18}.

All in all, the evolutionary picture that could be extracted from Figure~\ref{fig:SFQSB_active} is one in which the central BH and its host galaxy move around a similar \MBHMS\ scaling relation throughout their lifetime. They could start from a main-sequence or even starburst, gas-rich phase, evolving at an almost constant (specific) SFR, as also proposed by theoretical models \citep[e.g.][]{Lapi14, Aversa15} and direct observations \citep{2020A&A...642A..65C}, and then gradually switch off their accretion and star formation due to internal gas consumption, thus gradually reducing their SFR and accretion onto the central BH (right panel of Figure~\ref{fig:SFQSB_active}).  

\begin{figure*}
	\includegraphics[width=0.49\textwidth]{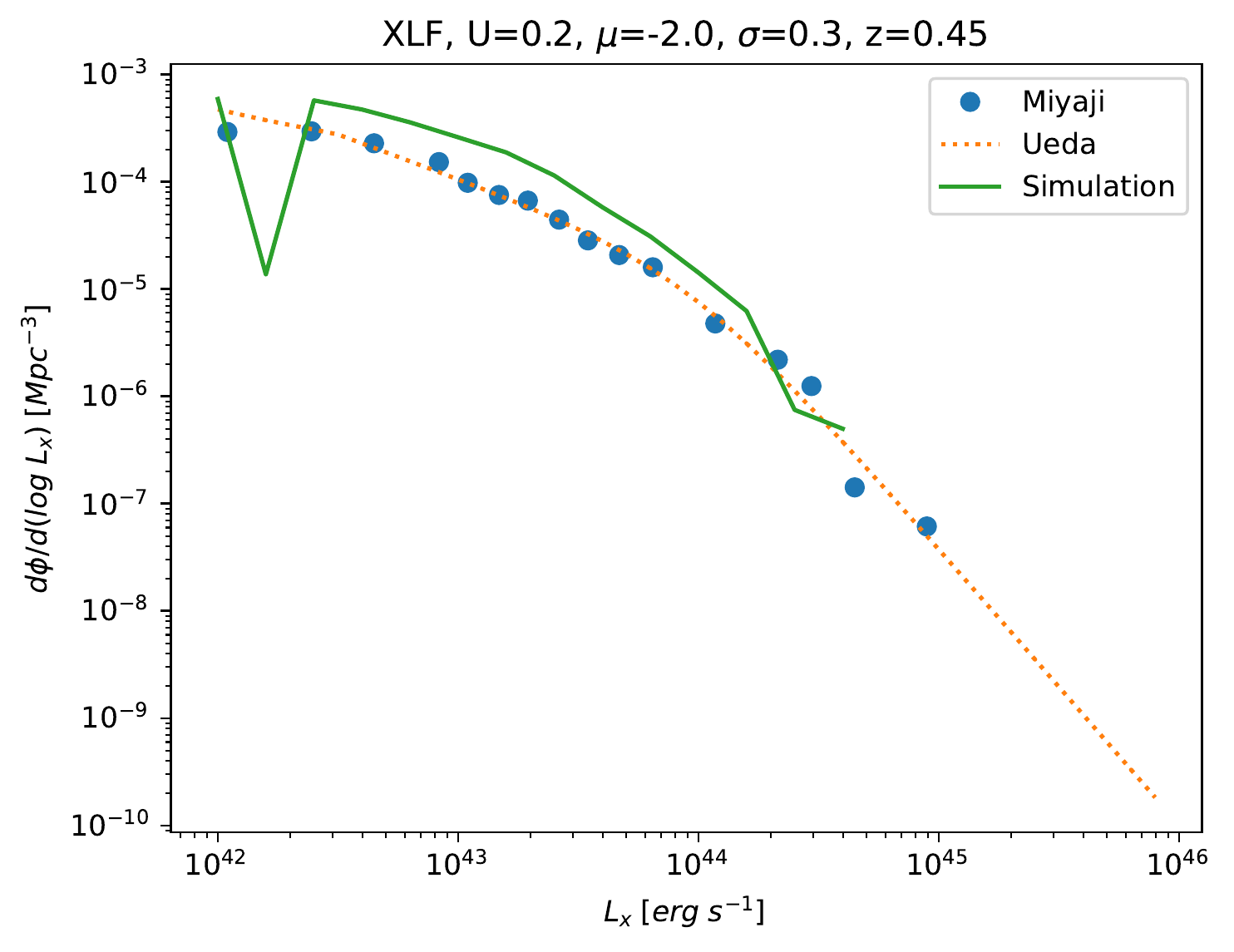}
	\includegraphics[width=0.49\textwidth]{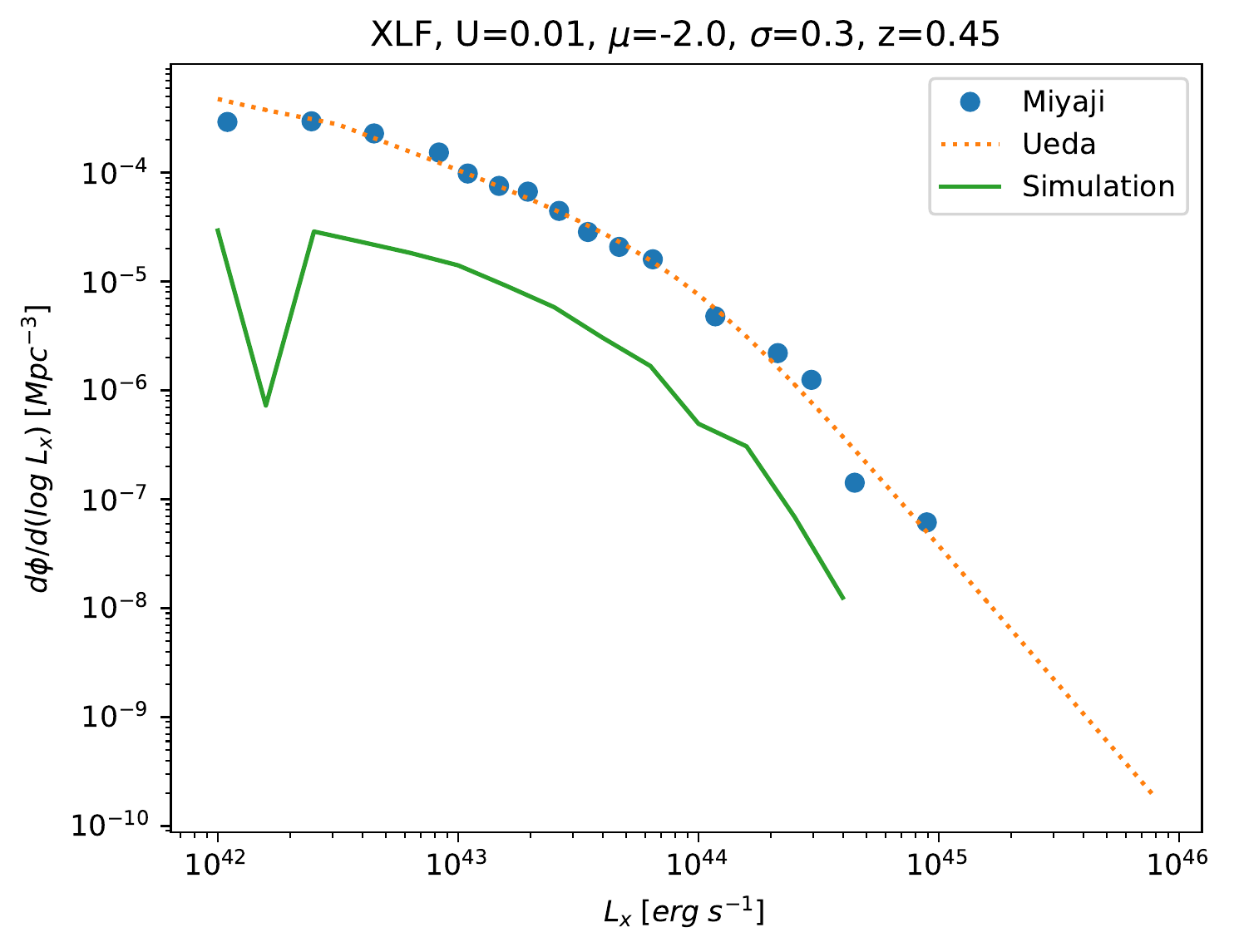}
	\includegraphics[width=0.49\textwidth]{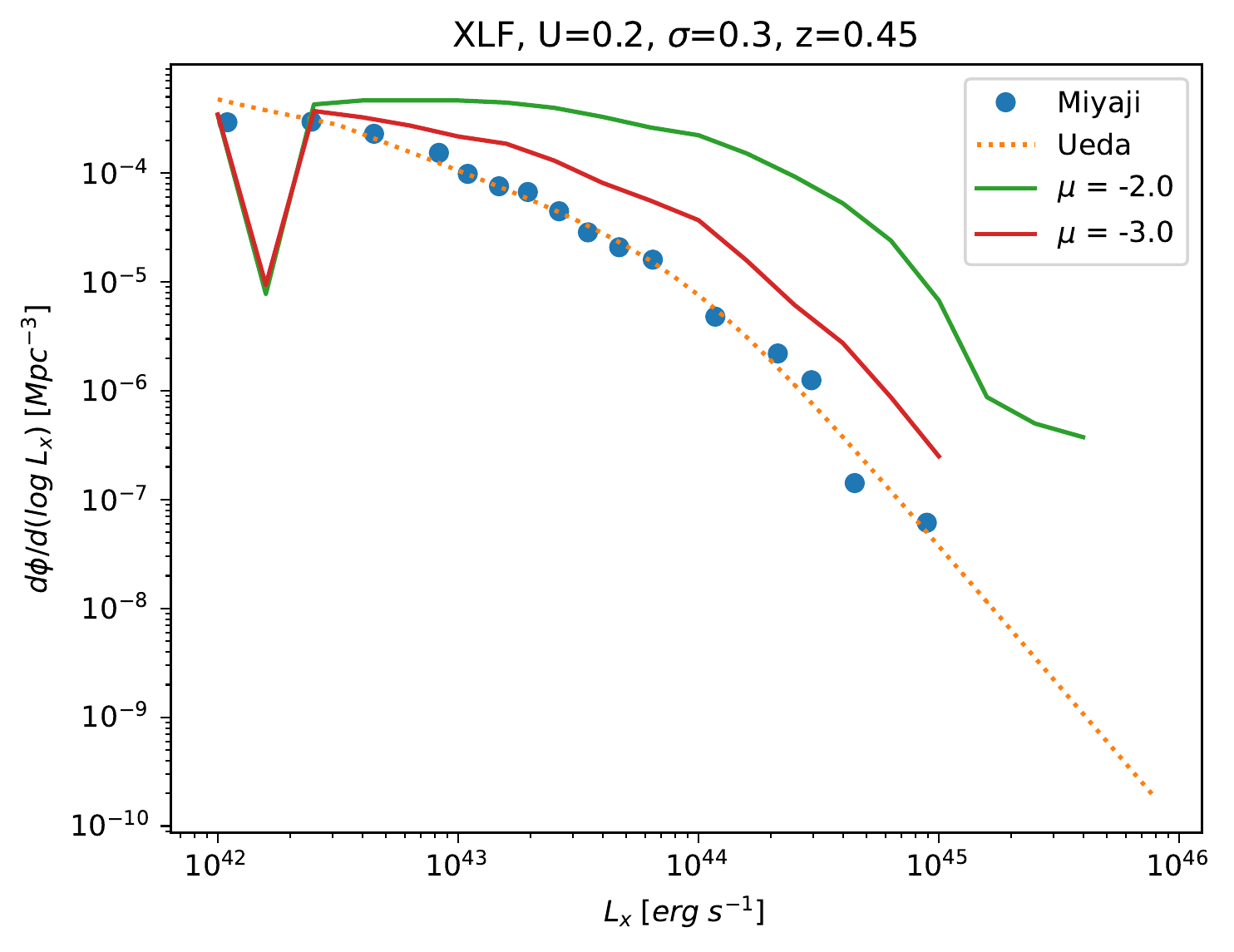}
	\includegraphics[width=0.49\textwidth]{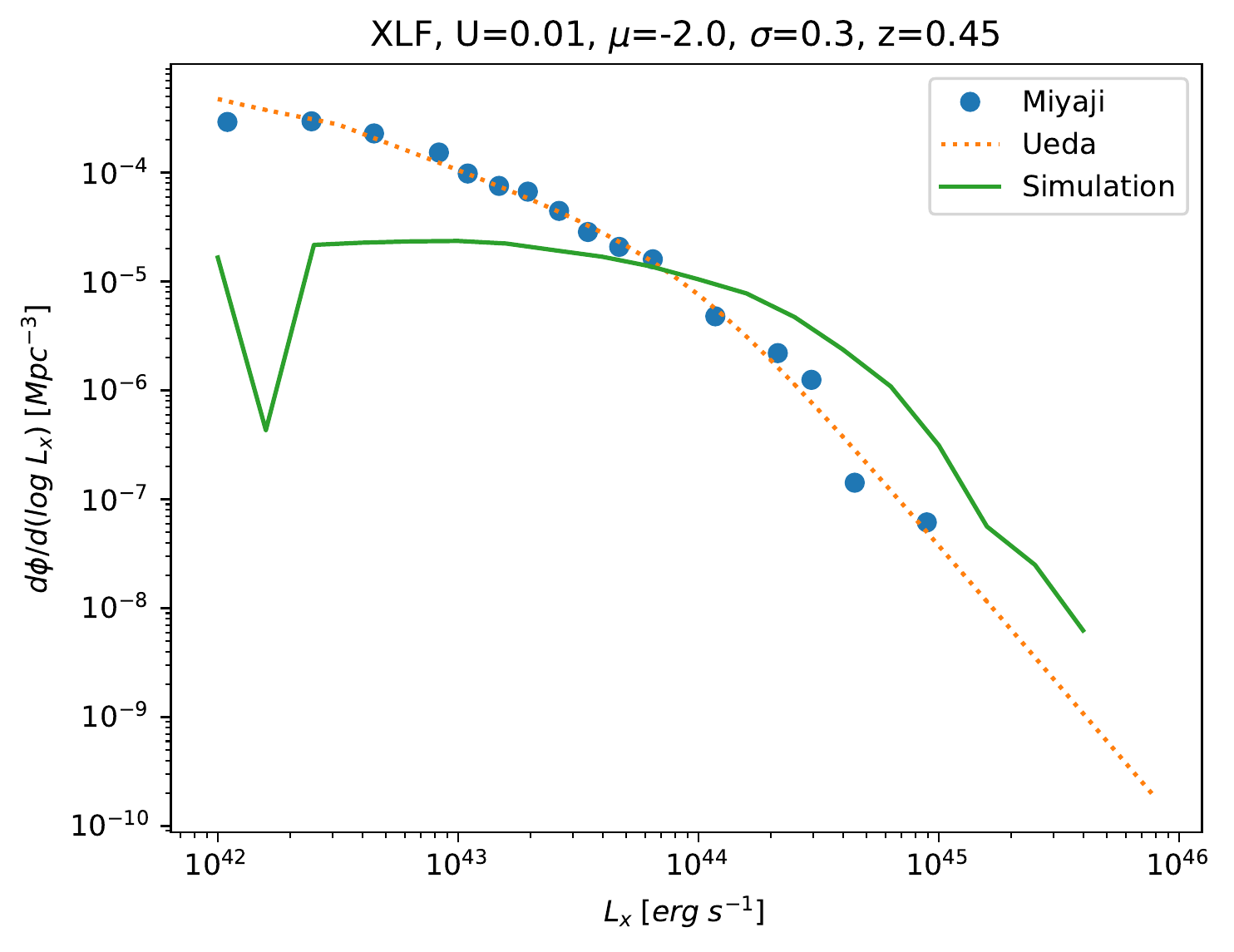}
	\caption{XLF from the models using the Eddington ratio better representing the $L_X$ of the data at $z=0.45$. Top panels: Using \citet{2015ApJ...813...82R} scaling relation, varying the duty cycle U=0.2 (left) and U=0.01 (right). Bottom panels: Using \citet{2019ApJ...876..155S} scaling relation, varying the duty cycle U=0.2 (left) and U=0.01 (right). Models are compared with data from \citet{2014ApJ...786..104U} and \citet{2015ApJ...804..104M} at the same redshift.}
	\label{fig:XLF}
\end{figure*}

\section{Discussion}\label{sec:disc}

We showed in the previous Sections that the mean \LXMS\ relation of X-ray detected active galaxies is a powerful tool to constrain the mean accretion rate of active BHs $\zeta_c$ as a function of time and BH mass, and in ways largely independent of the duty cycle. When coupled to other independent probes, the \LXMS\ can thus provide an invaluable support in breaking the degeneracies in the accretion parameters of supermassive BHs. 
For example, as discussed in Section 2, the AGN X-ray luminosity function is a convolution of the underlying BH mass function, which mostly depends on the BH-galaxy scaling relations \citep[e.g.,][]{Salucci99}, the intrinsic fraction of active BHs as a function of BH mass (the duty cycle $U(y,z)$), and the normalised Eddington ratio distribution \PLz\ \citep[see, e.g.,][and references therein]{Shankar13Acc}. Thus, knowledge of the AGN X-ray luminosity function and of the characteristic mean Eddington ratio $\zeta_c$ from independent observables, could shed light on the duty cycle, once a robust estimate of the underlying BH-galaxy scaling relation is available from, e.g., AGN clustering measurements \citep[see discussion in][]{ShankarNat,Allevato21,Viita21}. 

Figure~\ref{fig:XLF} shows a few examples of the dependencies of the AGN luminosity function on the most relevant model input parameters. We compare the observed X-ray AGN luminosity function\footnote{Both luminosity functions do not include Compton-thick sources, thus our duty cycle \UMBHz\ refers to the total fraction of Compton-thin AGN, i.e., those with $\log N_H<24\, {\rm cm^{-2}}$.} by \citet[][orange dotted lines]{2014ApJ...786..104U} and \citet[][blue filled circles]{2015ApJ...804..104M}, with the predictions of our reference model with a constant duty cycle $U=0.2$, a \MBHMS\ relation from \citet{2015ApJ...813...82R}, and a Gaussian \PLz\ with $\mu=-2$, a combination able to simultaneously reproduce the observed fraction of X-ray AGN (Figure~\ref{fig:AGN_fractions}) and mean \LXMS\ relation (Figure~\ref{fig:LX_M}). Despite the crudeness of our model, the top-left panel of Figure~\ref{fig:XLF} shows that our reference mock (solid green line) is able to broadly reproduce the data at all luminosities within a factor of $\lesssim 2$, without any extra fine-tuning. On the other hand, when switching to a \MBHMS\ relation with a higher normalisation than the one calibrated by \citet{2015ApJ...813...82R}, such as the one by \citet{2019ApJ...876..155S}, would tend to significantly overproduce the observed AGN luminosity function, an effect induced by the new \MBHMS\ relation which maps galaxies to more massive BHs and thus more luminous AGN \citep[e.g.,][]{ShankarNat}. To recover the match to the AGN luminosity function with the new \MBHMS\ relation we would require a mean Eddington ratio $\zeta_c$ significantly lower by more than an order of magnitude, as shown in the bottom, left panel (solid, red line), which allows to systematically shift the predicted luminosity function by a factor of $\gtrsim 10$ to fainter X-ray luminosities, in better agreement with the data. Although such a low value of $\zeta_c$ could still generate a \LXMS\ relation in broad agreement with the data, at least at larger stellar masses (by simply proportionally lowering the violet dashed model in the bottom, left panel of Figure~\ref{fig:LX_M}), and also with the observed AGN fraction (pink double dot-dashed line in the bottom right panel of Figure~\ref{fig:AGN_fractions}), it would be inconsistent with independent measurements of the mean Eddington ratios at similar redshifts \citep[e.g.,][]{H09,Kauff09,2019MNRAS.484.4360A}. Alternatively, we could keep the reference value of $\zeta_c$ but decrease the duty cycle to $U=0.01$, as shown in the solid lines reported in the right panels of Figure~\ref{fig:XLF}. This solution improves the match between the model with higher normalisation in the \MBHMS\ relation and the observed AGN luminosity function, at least at the bright end (bottom right panel). However, such a low value of the duty cycle $U=0.01$ is inconsistent with the much higher fraction of AGN detected in COSMOS-Legacy (Figure~\ref{fig:AGN_fractions}).

Our current work is able to provide additional clues and empirical evidence in support of the (complex) models of supermassive BH evolution in galaxies. According to the standard picture of the early phases of the co-evolution of galaxies and their central BHs \citep[e.g.,][]{Granato06,Hopkins06,Lapi18}, galaxies undergo a first rapid, gas-rich and strong burst of star formation, during which a (seed) BH can substantially grow at or above the Eddington limit, followed by a more regular and then quiescent phase during which both the star formation and the accretion onto the central BH drop substantially. We already showed in the left panel of Figure~\ref{fig:SFQSB_active} that, in the context of our modelling, when assuming a constant or slowly varying underlying \MBHMS scaling relation, the data tend to favour an evolving characteristic Eddington ratio $\zeta_c$, steadily declining when the galaxy transitions from the starburst to the quiescent phase, and we suggested, based on the comparison with the L$_X$-SFR relation (right panel of Figure~\ref{fig:SFQSB_active}), that this temporal trend in BH accretion rate should be closely mirrored by the star formation in the host galaxy, in agreement with the expectations from theoretical models. Here we further elaborate on this idea.
In our previous work \citep[see, e.g., Figure~3 in][]{2020A&A...642A..65C}, we showed that main-sequence and quiescent galaxies share similar ratios of BHAR and SFR at all probed cosmic epochs, suggesting that the two processes are indeed linked together throughout different galaxy phases.  
In fact, the mean BHAR/SFR can be written as ${\rm BHAR/SFR} \propto L_{bol}/{\rm SFR} \propto 10^{\zeta_c} {\rm M}_{\rm BH}/(k {\rm M}_*)$, where $k={\rm SFR}/{\rm M}_*$ is the specific SFR. Thus, at fixed ${\rm M}_{\rm BH}/{\rm M}_*$, a similar BHAR/SFR ratio as the one observed in star forming and quiescent galaxies, would be induced by a proportional decline in characteristic Eddington ratio $\zeta_c$ and specific SFR $k$ within a bin of stellar mass. Analogously, the significantly lower BHAR/SFR in starbursts with respect to quiescent/star forming galaxies, as measured by \citet[][]{2020A&A...642A..65C}, would be naturally interpreted as a proportionally higher specific SFR $k$ 
and roughly constant or slightly higher $\zeta_c$ in these young gas rich systems, as predicted by some BH evolutionary models \citep[e.g.,][]{Lapi14,Aversa15}.

\section{Conclusions}\label{sec:concl}

In this work we use statistical semi-empirical models to generate accurate mock catalogues of active galaxies, which we analyse in the same manner as in the comparison observational sample from \citet{2020A&A...642A..65C}. Our goal is to unveil the input parameters driving the \LXMS\ relation. We start from a halo mass function at a given redshift, we assign galaxies and BHs to dark matter haloes via the most up-to-date empirical stellar-halo and \MBHMS\ relations, and we assume a SFR depending only on stellar mass and redshift. We explore a range of Eddington ratio distributions \PLz, \MBHMS\ scaling relations and duty cycles \UMBHz. Our results can be summarised as follows:
\begin{itemize}
    \item In agreement with previous findings \citep[see, e.g.,][]{2012ApJ...746...90A,Shankar13Acc}, the apparent increase of AGN detections towards high stellar masses, i.e., the ``observed'' AGN fraction, is not necessarily caused by AGN being more frequent in more massive galaxies, but we find that it is mostly a consequence of the X-ray survey flux limit, which prevents the detection of the faintest sources with a higher probability of being located in lower mass galaxies.
    \item The mean \LXMS\ (or L$_{\rm X}$-SFR) relation in detected BHs is largely independent of the AGN duty cycle, but strongly depends on the {shape, normalisation and scatter of the underlying \MBHMS\ scaling relation} and on the characteristic Eddington ratio $\zeta_c$, which play a degenerate role in linking the mean \LX\ with the BH mass. 
    \item When assuming a roughly constant \MBHMS\ relation with time, as indicated by many recent observations, current X-ray data on the \LXMS\ relation favour models with a mean Eddington ratio of a few percent at $z=0.45$ and rapidly approaching the Eddington limit at $z\sim 3$, in broad agreement with a variety of independent data sets and theoretical models. 
    \item At fixed redshift $z \gtrsim 1$, the same data sets also show evidence for downsizing, with the most massive BHs having accreted their mass more rapidly than less massive BHs.  
    \item At fixed redshift, the \LXMS\ relation increases by nearly an order of magnitude in normalisation when moving from quiescent to starburst galaxies. Our models suggest that, on the reasonable assumption of a constant \MBHMS\ relation, this increase in mean \LX\ is mostly induced by the mean $\zeta_c$ being much higher during the starburst, gas-rich phase, and rapidly dropping in the quiescent, gas-poor phase. 
    \item Models consistent with the observed \LXMS\ relation, independent measurements of the mean Eddington ratios, the observed X-ray AGN fraction, and the X-ray AGN luminosity function, are characterised by input \MBHMS\ relations with normalizations aligned with those of local AGN samples \citep[e.g.,][]{2015ApJ...813...82R,Shankar19}, which are often lower than those derived from dynamically measured local BHs.
\end{itemize}
The main result derived from this work is the evidence that the \LXMS\ relation can efficiently break degeneracies among input duty cycles, Eddington ratio distributions and also BH-galaxy scaling relations, when the latter are coupled with independent observational probes, such as AGN clustering measurements \citep{ShankarNat} and observed AGN fractions, thus representing a powerful test for BH evolutionary models in a cosmological context.

\section*{Acknowledgements}
We warmly thank the anonymous referee for an extremely careful reading of the manuscript and for a number of excellent inputs that have significantly improved the analysis and presentation of the results.
We also acknowledge Guang Yang, James Mullaney and Emanuele Daddi for useful conversations.
RC acknowledges financial support from CONICYT Doctorado Nacional 21161487, CONICYT PIA ACT172033. 
and the Max-Planck Society through a Partner Group grant with MPA. 
FS acknowledges partial support from a Leverhulme Trust Research Fellowship. 
AL is partially supported by the PRIN MIUR 2017 prot. 20173ML3WW 002,
`Opening the ALMA window on the cosmic evolution of gas, stars and supermassive black holes',
by the MIUR grant `Finanziamento annuale individuale attivit\'a base di ricerca' and by
the EU H2020-MSCA-ITN-2019 Project 860744 `BiD4BEST:
Big Data applications for Black hole Evolution STudies'.
FS and AL also acknowledge partial support from the EU H2020-MSCA-ITN-2019 Project 860744. VA acknowledges support from INAF-PRIN 1.05.01.85.08.

Software: NumPy \citep{2020NumPy-Array}, SciPy \citep{2020SciPy-NMeth}, pandas \citep{reback2020pandas}, Matplotlib \citep{Hunter:2007}, COLOSSUS Python package \citep{2018ApJS..239...35D}.


\section*{Data Availability}
The data underlying this article will be shared on reasonable request to the corresponding author.



\bibliographystyle{mnras}
\bibliography{bibliography} 







\bsp	
\label{lastpage}
\end{document}